\documentclass[journal,onecolumn]{IEEEtran}
\IEEEoverridecommandlockouts

\usepackage[latin1]{inputenc}
\usepackage[cmex10]{amsmath}
\usepackage{amsfonts}
\usepackage{amssymb}
\usepackage{threeparttable}
\usepackage{soul}

\usepackage{verbatim}
\usepackage{array}
\usepackage{multirow}
\usepackage{dcolumn}
\usepackage{xcolor}
\usepackage{url}
\usepackage{graphicx}
\DeclareGraphicsExtensions{.eps}

\usepackage{bbm}
\usepackage{float} 

\usepackage{mathrsfs}
\usepackage{color}
\usepackage{booktabs}

\begin{document}

\title{Alignment-free characterization of polarizing beamsplitters}

\author{Felipe Calliari, Pedro Tovar, Christiano Nascimento, Breno Perlingeiro, Gustavo Amaral, and Guilherme Tempor{\~a}o %
\thanks{}
\thanks{F. Calliari, P. Tovar, C. Nascimento, B. Perlingeiro and G. Tempor{\~a}o are with the Center for Telecommunications Studies, Pontifical Catholic University of Rio de Janeiro (CETUC/PUC-Rio), Rio de Janeiro, RJ, Brazil (e-mail: \{felipe.calliari, ptovar, cnascimento, bperlingeiro\}@opto.cetuc.puc-rio.br and temporao@puc-rio.br).}
\thanks{G.~C.~Amaral is with the Center for Telecommunications Studies, Pontifical Catholic University of Rio de Janeiro, RJ, Brazil and with the QC2DLab, Kavli Foundation, Technical University of Delft, The Netherlands (e-mail: gustavo@opto.cetuc.puc-rio.br).}
}

\maketitle
\begin{abstract}
Traditional methods for measurement of Polarizing Beamsplitter (PBS) parameters, especially the extinction ratio, require highly polarized light sources, alignment procedures and/or experimental parameters that change over time, such as polarization rotations. In this work, a new method is presented, which employs unpolarized light and a Faraday Mirror. It is shown that precise extinction ratio and insertion loss values can be achieved in three single-sweep measurements, without any alignment requirements or time-varying signals of any kind.
\end{abstract}

\section{Introduction}
\label{sec:intro}
Polarizing Beamsplitters (PBS) have been employed in a plethora of applications in both free-space and fiber optical classical and quantum communications \cite{lo2012measurement, kaiser2012entanglement, amaral2016time, tyan1997design}. They are often used to split unpolarized light into two polarized components, transmitting p-polarized light while reflecting s-polarized light. One of the most important parameters for a PBS -- or any polarizer -- is the extinction ratio, which measures the leakage fraction of s-polarized light into the p-polarization output mode and vice-versa.  

Traditionally, this measurement is performed using a highly coherent light source, such as a laser \cite{tyan1997design, saleh1991fundamentals, penninckx2005definition, abadia2017novel, zhang2016high, xu2016compact, elfiky2016high, zaoui2010photonic}. The polarization state of a test light signal is matched with both axes -- horizontal and vertical polarizations -- of the device under test (DUT), using birefringent elements such as half-waveplates or fiber polarization controllers, depending on whether the DUT is a bulk or fiber-optical PBS, respectively. This method, however, has two main drawbacks: the requirement of alignment, especially in the case of fiber-optical devices, where there is no obvious reference for what a ''horizontal'' polarization state should be; and the requirement of the test light signal to be highly polarized, i.e., the ratio between the power in any two orthogonal polarization components must be at least one order of magnitude higher than the extinction ratio of the DUT.

For the first limitation, a time-varying birefringence, such as a polarization controller, is usually employed in order to search for the input polarization state that minimizes one of the outputs. For the second, due to the unpolarized nature of amplified spontaneous emission (ASE) of any laser, care must be taken to appropriately filter out sidebands that can potentially contain unpolarized light \cite{derickson1998fiber}. Alternatively, any light source followed by a polarizer can be employed, as long as the polarizer's extinction ratio is, again, at least one order of magnitude higher than the DUT's.

In this paper, we present a novel technique for measuring the extinction ratio of any 3- or 4-port PBS that has no alignment requirements and that employs an unpolarized light source, such as a LED. By a clever arrangement of a Faraday Mirror, a 2x2 optical switch, two power meters and an optical circulator, it is possible to obtain the insertion losses and extinction ratios in a single-sweep, three-step measurement. The method is validated, a fully automated characterization unit is assembled and the results are compared to the literature method \cite{tyan1997design,saleh1991fundamentals,penninckx2005definition,abadia2017novel} with good correspondence.

\section{Method}
\label{sec:method}

\begin{figure}[h!]
\centering\includegraphics[width=8cm]{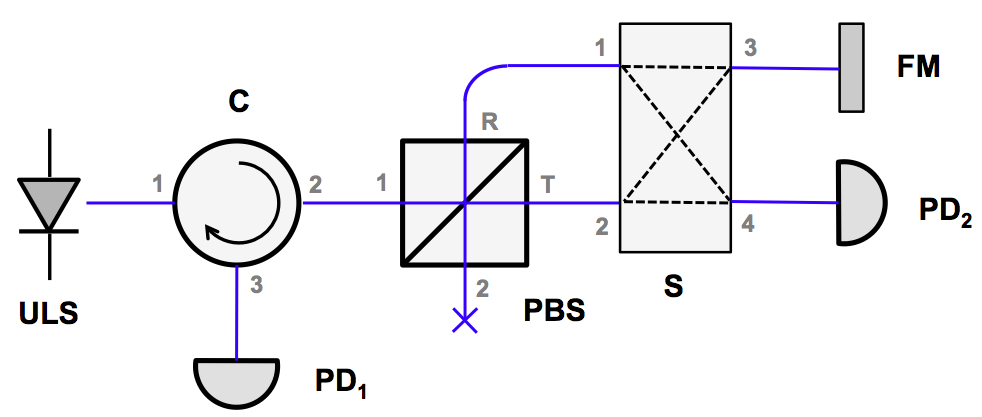}
\caption{Basic scheme for characterization of a PBS. ULS: unpolarized light source; C: optical circulator; PD: photodetector; PBS: polarizing beamsplitter; S: 2x2 optical switch; FM: Faraday mirror. Port 2 of the PBS is connected to an optical termination.}
\label{fig:scheme}
\end{figure}

The proposed scheme for determining the insertion losses and extinction ratios of a PBS involves two steps, as shown in Fig. \ref{fig:scheme}. Each step corresponds to a state of the optical switch (S). In the first step, a connection is established between ports 1-3 and 2-4 of S, whereas in the second step the connections are switched to 1-4 and 2-3.

Let us assume for the moment that the polarization dependent loss (PDL) of all other components are negligible; we will return to this point later in section \ref{sec:discussions}. This means that the polarization state in port 1 of the PBS is a completely mixed state, generated by the unpolarized light source (ULS). Let the insertion losses of the PBS between ports 1 and $j$ be represented by $\alpha_{i}^{j}$, where the subscript $i$ refers to the horizontal (H) or vertical (V) polarization state, assumed to coincide with the PBS eigenstates. We define the \textit{extinction ratios} (ER) of ports \textbf{R} and \textbf{T}, respectively, as:
\begin{equation}
\label{eq:1}
\xi_R = {\alpha_V^{R}}/{\alpha_H^{R}}
\end{equation}
\begin{equation}
\label{eq:2}
\xi_T = {\alpha_H^{T}}/{\alpha_V^{T}}.
\end{equation}
Note that a device is not characterized by a single ER value; rather, the ER is measured for combinations of its input and output modes. In fact, many devices exhibit differences between $\xi_R$ and $\xi_T$ that can be over an order of magnitude. 

The goal of the method is finding the values of the polarization-dependent insertion losses and, for simplicity, we will assume henceforth that 3-port devices are being considered. In the case of a 4-port PBS, we always assume the unused port is connected to an optical termination that prevents spurious back-reflections from influencing the measurements. In this case, the method must be applied a second time, by connecting the circulator to PBS port 2 and the termination to port 1.

Let $P_0$ be the input power in the DUT port 1. In the first step of the measurement, the transmitted power is directly detected by PD$_2$, whereas the reflected light impinges on the FM and is reflected back to port 1 and then to PD$_1$ via the circulator. The action of the FM is such that the insertion losses in the back propagation are swapped with respect to the input ones, such that the powers $P_1$ and $P_2$ in the photodetectors are given by:
\begin{equation}
\label{eq:3}
P_1 = \ell^C_{23}\left( \ell^S_{13} \right)^2\ell^{FM}\alpha_H^{R}\alpha_V^R \frac{P_0}{2}
\end{equation}
\begin{equation}
\label{eq:4}
P_2 = \ell^S_{24}\left(\alpha_H^{T}+\alpha_V^T \right) \frac{P_0}{2},
\end{equation}
where $\ell^C_{ij}$, $\ell^S_{ij}$ and $\ell^{FM}$ represent the losses between ports $i$ and $j$ of the circulator and optical switch, and the insertion loss of the FM, respectively. Note that the FM insertion loss corresponds to the total loss experienced by the light upon reflection at the device, thereby appearing in the equation as a linear (non-squared) term. Also note that any polarization rotations in the fiber pigtails connecting the components are irrelevant, due to the FM presence. It is evident that Eqs. (\ref{eq:3}) and (\ref{eq:4}) are insufficient for determining the insertion losses. The additional equations are obtained in the second step of the procedure, changing the switch connections. Now, the FM is connected to port T and photodetector D$_2$ to port R of the PBS. The new readings in the photodetectors are given by:
\begin{equation}
\label{eq:5}
P_1' = \ell^C_{23}\left( \ell^S_{23} \right)^2\ell^{FM}\alpha_H^{T}\alpha_V^T \frac{P_0}{2}
\end{equation}
\begin{equation}
\label{eq:6}
P_2' = \ell^S_{14}\left(\alpha_H^{R}+\alpha_V^R \right) \frac{P_0}{2}.
\end{equation}

Note that we now have two pairs of equations: Eqs. (\ref{eq:3}, \ref{eq:6}) for the reflection coefficients $\alpha_{H,V}^R$ and Eqs. (\ref{eq:4},\ref{eq:5}) for the transmission coefficients $\alpha_{H,V}^T$. Letting $(p_1,p_2,p_1',p_2')$ be the normalized versions of the photodetector readings with respect to the input power $P_0$, we obtain:
\begin{equation}
\label{eq:7}
\alpha_{H,V}^T = \frac{p_2}{\ell_{24}^S}\pm\sqrt{\left(\frac{p_2}{\ell_{24}^S}\right)^2-\frac{2p_1'}{\ell^C_{23}\left( \ell^S_{23}\right)^2\ell^{FM}}}
\end{equation}
\begin{equation}
\label{eq:8}
\alpha_{H,V}^R = \frac{p_2'}{\ell_{14}^S}\mp\sqrt{\left(\frac{p_2'}{\ell_{14}^S}\right)^2-\frac{2p_1}{\ell^C_{23}\left( \ell^S_{13}\right)^2\ell^{FM}}},
\end{equation}
where the upper and lower sign correspond to $H$ and $V$ polarizations, respectively. The signs were assigned by previous knowledge of which polarization coefficients are higher in each output. It is clear that knowing all losses in the circulator and optical switch allows for the transmission coefficients to be precisely calculated. In practice, the powers ($P_1,P_1'$) are orders of magnitude lower than ($P_2,P_2'$), such that the following first-order approximations hold:
\begin{align}
\begin{aligned}
\alpha_H^T &\approx 2p_2/\ell_{24}^S ;\\
\alpha_V^T &\approx p_1'\ell_{24}^S/p_2\ell_{23}^C\left( \ell^S_{23}\right)^2 \ell^{FM} ;\\
\alpha_V^R &\approx 2p_2'/\ell_{14}^S ;\\
\alpha_H^R &\approx p_1\ell_{14}^S/p_{2}'\ell_{23}^C\left( \ell^S_{13}\right)^2\ell^{FM}.
\end{aligned}
\end{align}
Therefore, the ERs as defined by Eqs. (\ref{eq:1}) and (\ref{eq:2}) can be approximately written, in decibels, as:

\begin{equation}
\label{eq:9}
\xi_R = 10\,\text{log}\left(\frac{2\ell_{23}^C\left( \ell^S_{13}\right)^2\ell^{FM}}{\ell_{14}^S\ell_{14}^S}\frac{P_2'^2}{P_1 P_0}\right)
\end{equation}
\begin{equation}
\label{eq:10}
\xi_T = 10\,\text{log}\left(\frac{2\ell_{23}^C\left( \ell^S_{23}\right)^2\ell^{FM}}{\ell_{24}^S\ell_{24}^S}\frac{P_2^2}{P_1' P_0}\right),
\end{equation}
which clearly shows that the extinction ratios can be easily obtained from the power measurements $P_1$, $P_1'$, $P_2$, $P_2'$ and $P_0$.

\section{Experimental results}
\label{sec:setup}

\begin{figure}[h!]
\centering\includegraphics[width=12cm]{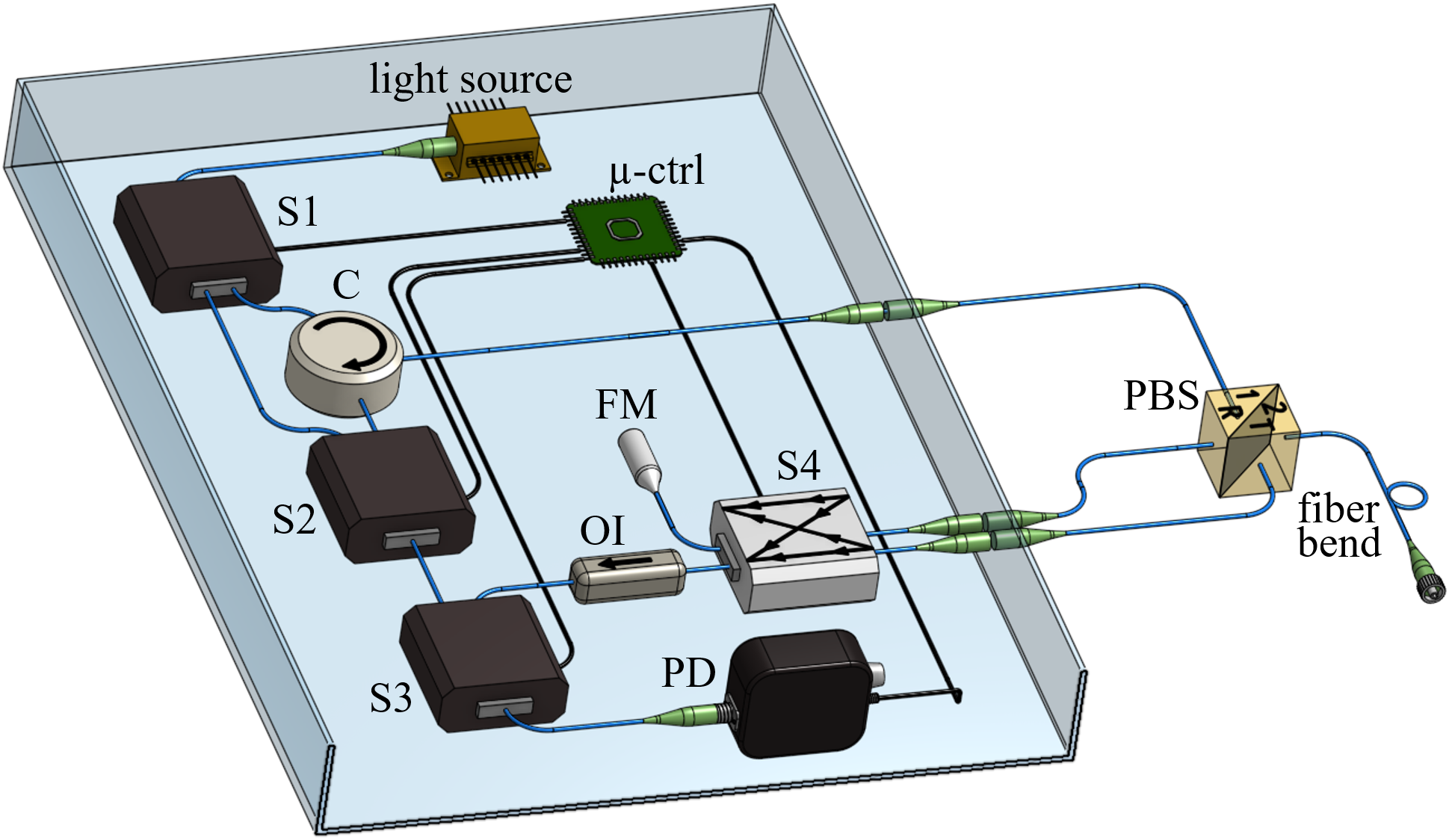}
\caption{Experimental setup for the extinction ratio measurement of polarizing beamsplitters. C: optical circulator; FM: Faraday mirror; OI: optical isolator; PBS: polarizing beamsplitter; PD: photodetector; S1-3: 2x1 optical switch; S4: 2x2 optical switch.}
\label{fig:exp_setup}
\end{figure}

Fig. \ref{fig:exp_setup} depicts the experimental setup for a fully automated characterization of the polarizing beamsplitter. In this particular setup, only one photodetector (PD) is employed, at the expense of using more optical switches. A microcontroller is used to operate four optical switches and to measure the different optical powers. The probe light, an unpolarized light source, is comprised of a light emitting diode (LED) followed by a 3-nm wide optical band-pass filter; this is to ensure that the bandwidth of the probe light falls within the smallest bandwidth of the employed devices. The limiting device, in this case, is the Faraday Mirror, with a 17 nm bandwidth centered around 1550 nm.

First, the probe light is connected to the PD by switching S1, S2, and S3. The optical input power in the DUT ($P_0$) is obtained by the measured optical power and the losses in the optical switches and the circulator, which are previously characterized. The voltage value in the PD is then converted into a digital value by an Analog-to-Digital Converter (ADC) and then interpreted by the microcontroller. Second, changing the states of S1 and S2, the PD measures the reflected (port R) optical power from the circulator which corresponds, in Section \ref{sec:method}, to $P_1$. In the third step, by changing the state of S3, the transmitted optical power (PBS port T) corresponding to $P_2$ is measured after passing through the 2x2 switch (S4) \cite{22switch} and the optical isolator whose losses are also previously characterized. Finally, by changing the states of S3 and S4, the microcontroller repeats the second and third steps, obtaining the returning transmitted optical power from the circulator, $P_1'$, and the reflected optical power, $P_2'$.

For the characterization of a 3-port PBS, the steps presented above are sufficient. However, in order to characterize a 4-port PBS, the leftover port should be terminated in order to extinguish any spurious reflected light. It is important to note that the equations used in the microcontroller are slightly different from those shown in Section \ref{sec:method}, since the assembly is different. The main difference is the use of a single PD, in addition to three optical switches (S1, S2 and S3) and an optical isolator, which were included to allow for a fully automated measurement. Note that an automated measurement is only possible for a 3-port PBS, as a 4-port device requires a connection change between ports 1 and 2 for full characterization. Considering the additional devices' losses and following the same steps of Section \ref{sec:method}, the ERs can be written as:

\begin{equation}
\label{eq:12}
\xi_{i,R} = 10\,\text{log}\left(\frac{2 \left( \ell_{13}^{S4}\right)^2\ell^{FM} \ell_{23}^C \ell_{12}^{S2} \ell_{13}^{S3} }{ \ell_{12}^{S1} \ell_{12}^{C} \left( \ell_{14}^{S4} \ell^{OI} \ell_{12}^{S3} \right)^2 }\frac{P_2'^2}{P_1 P_0}\right), \qquad \qquad i \in [1,2]
\end{equation}

\begin{equation}
\label{eq:13}
\xi_{i,T} = 10\,\text{log}\left(\frac{2 \left( \ell_{23}^{S4}\right)^2\ell^{FM} \ell_{23}^{C} \ell_{12}^{S2} \ell_{13}^{S3} }{ \ell_{12}^{S1} \ell_{12}^{C} \left( \ell_{24}^{S4} \ell^{OI} \ell_{12}^{S3} \right)^2 }\frac{P_2^2}{P_1' P_0}\right), \qquad \qquad i \in [1,2]
\end{equation}

Two 3-port and one 4-port PBS's were characterized with the proposed method, and the results are presented in Table \ref{tab:er_pbs}. Even though all extinction ratios measured comply with those described in the devices' data sheets (> 20 dB), for a more accurate comparison, the extinction ratios were also measured with the conventional method described in literature \cite{tyan1997design,saleh1991fundamentals,penninckx2005definition,abadia2017novel}. For this, a polarized light source comprised of a laser was used and had its polarization aligned with one axis of the tested PBS. For precise polarization alignment, a polarization tracker followed by a mechanical polarization controller (PC) were included, such that the tracker provides rough polarization adjustment, whereas the mechanical controller grants fine tuning, thus maximizing the power transmission for the aligned port. Assuming that the degree of polarization (DOP) of the polarized light source and the polarization alignment accuracy are not limiting factors, the optical power ratio between the measured powers of either output ports corresponds to the ER of the input-output pair. Table \ref{tab:er_pbs} shows the measurement results for both methods. The so-called literature method is the one described in \cite{saleh1991fundamentals,abadia2017novel}.

\begin{table}[htb!]
\centering
\begin{threeparttable}[b]
\caption{Experimental results: measurements of Extinction Ratios for different combinations of input and output modes in dB.}
\label{tab:er_pbs}
\begin{tabular}{lllllllllll}
\hline
    & \multicolumn{2}{c}{3-Port PBS 1} & & \multicolumn{2}{c}{3-Port PBS 2} & & \multicolumn{4}{c}{4-Port PBS} \\ \cline{2-3} \cline{5-6} \cline{8-11}
    & \multicolumn{1}{c}{1-T} & \multicolumn{1}{c}{1-R} & & \multicolumn{1}{c}{1-T} & \multicolumn{1}{c}{1-R} & & \multicolumn{1}{c}{1-T} & \multicolumn{1}{c}{1-R} & \multicolumn{1}{c}{2-T} & \multicolumn{1}{c}{2-R}\\ \hline
\multicolumn{1}{r}{Literature method} & \multicolumn{1}{c}{30.8} & \multicolumn{1}{c}{31.0} & & \multicolumn{1}{c}{27.5} & \multicolumn{1}{c}{25.0} & & \multicolumn{1}{c}{24.8} & \multicolumn{1}{c}{23.8} & \multicolumn{1}{c}{23.6} & \multicolumn{1}{c}{25.1}\\
\multicolumn{1}{r}{Proposed method} & \multicolumn{1}{c}{31.6} & \multicolumn{1}{c}{31.9} & & \multicolumn{1}{c}{27.6} & \multicolumn{1}{c}{26.7} & & \multicolumn{1}{c}{28.2} & \multicolumn{1}{c}{25.7} & \multicolumn{1}{c}{22.9} & \multicolumn{1}{c}{28.8}\\

\hline
\end{tabular}
\end{threeparttable}
\end{table}

\begin{figure}[h!]
\centering\includegraphics[width=8cm]{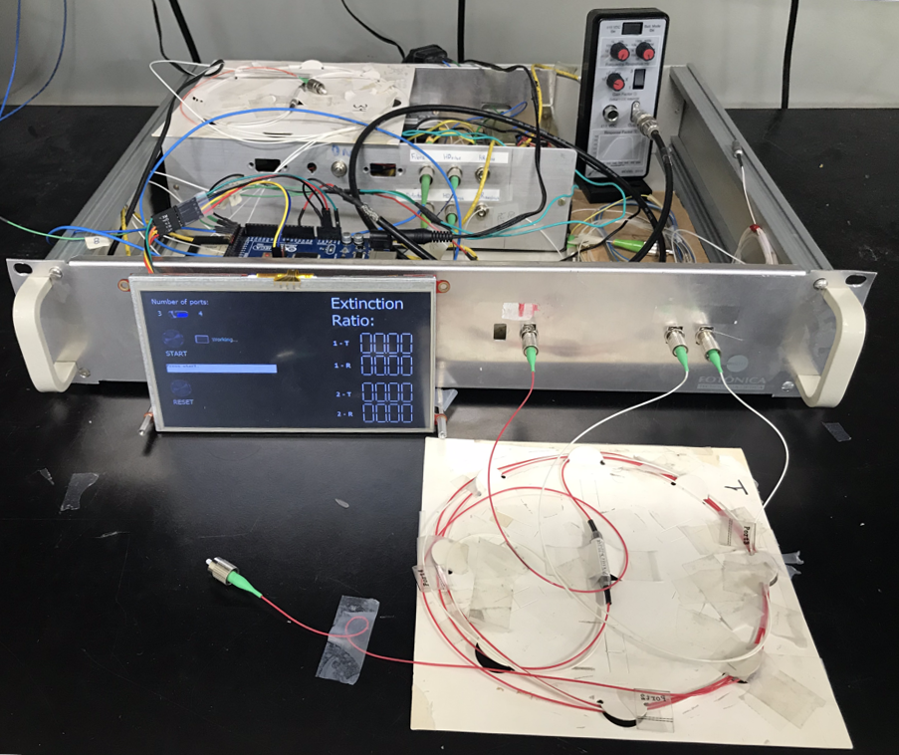}
\caption{Photo of the experimental setup depicting the measurement unit as well as its user interface and the DUT.}
\label{fig:Photo_setup}
\end{figure}

According to Table \ref{tab:er_pbs},the proposed and literature methods provide similar extinction ratio results which prove the efficacy of the proposed method. The difference between results is attributed to the usage of a polarized source in the literature method, and to the fact that it has to be precisely aligned to the axis of the PBS in order for the device to be accurately characterized. The proposed method, on the other hand, does not introduce any operational errors since it does not require any kind of alignment. In order to propitiate a practical handling of the system, an interface was implemented with a user friendly front panel, which is shown in  Fig. \ref{fig:Photo_setup}. The microcontroller allows for the user to input the number of ports and to control the measurement with start and reset buttons.

\section{Discussions}
\label{sec:discussions}

\subsection{Polarization dependent losses of the components}
\label{sec:pdl}

In section \ref{sec:method}, we have ignored the effect of the PDL of the circulator and optical switch, for simplicity's sake. It is well known, however, that these devices have non-negligible PDL, and therefore both should be considered in the calculations.

By assuming that the switch insertion losses $\ell_{ij}^S$ are polarization-dependent, one can write them as $\ell_{ij}^H$ and $\ell_{ij}^V$ (the "S" superscript has been dropped for simplicity). First, the analysis of the effect of the back-reflected light by the Faraday mirror is performed. Note that, regardless of the polarization state at the switch's input (say, in port 1), the orthogonal polarization components will be submitted to the same round-trip loss. This is due to the effect of the FM, which swaps the polarization components, such that the power fraction that suffered an attenuation $\ell_{1,3}^H$ in the forward direction, will suffer an attenuation $\ell_{1,3}^V$ in the backward direction. Thus, the round-trip loss is given by $\ell_{1,3}^H\ell_{1,3}^{V}$ for any polarization state, and the power detected by PD$_1$ is unaffected by the PDL of the optical switch.

The transmission effect of the optical switch, on the other hand, is polarization-dependent. As the fiber patch cords connecting the PBS outputs to the optical switch introduce random polarization rotations, the input polarization states in both of its ports are unknown. Therefore, an uncertainty upper bounded by the optical switch's PDL is added to the power value measured by PD$_2$ where, typically, values in the range 0.1-0.3 dB are expected. The exception here are PBS's with polarization-maintaining fibers, where the polarization state is kept unaltered. In this case, a simple polarization controller can be employed to avoid the measurement uncertainty. 

Concerning the circulator, its PDL effects can indeed be circumvented if information about its axis and magnitude is available. From the point of view of PD$_2$, the state of polarization of the light source could be chosen such that it pre-compensates the effect of the circulator's PDL, in which case, after going through it, the DOP is effectively zero. From the point of view of PD$_1$, on the other hand, the joint effect of the PDL from the 1-2 and 2-3 ports needs to be neutralized, which could be accomplished by adding an extra PDL element immediately before PD$_1$, in such a way that the light impinging on the detector is, again, unpolarized. It should be noted that this pre-compensation scheme is utterly impractical and the uncertainties introduced in the measurements due to lack of knowledge about the circulator's PDL are extremely low.

\subsection{Degree of Polarization of the Light Source}
\label{sec:dop}

Even though creating light with low DOP can be easily done in practice - such as with a laser and either a high-frequency polarization scrambler or two orthogonally spliced polarization maintaining fibers, or simply an LED, as has been done in this experiment - it is interesting to analyze the effects of a nonzero DOP. Not only can one identify DOP intervals within which the method is still valid, but also determine the amount of uncertainty arising from the worst case scenario.

Both detected powers $P_1$ and $P_2$ would be affected by the presence of partially-polarized light. Let $P_0 = (1-\text{DOP})\cdot P_u + \text{DOP}\cdot P_p$, where $P_u$ and $P_p$ correspond to the unpolarized and polarized portions of the input power, respectively. If the polarization state of the polarized portion is aligned such that half of its power is directed to both outputs of the PBS (i.e., a linear combination of the PBS's eigenstates with equal probability amplitudes), then no error will arise. However, such solution would defeat the purpose of an alignment-free method.

In the worst case scenario, the polarized portion would be aligned to one of the PBS's eigenstates. Then, the term $P_0/2$ in eqs. (\ref{eq:4}) and (\ref{eq:6}) would now read $P_u/2 + P_p$ or $P_u/2$, depending on whether the polarized portion is aligned to the R or T output mode, whereas in eqs. (\ref{eq:3}) and (\ref{eq:5}) $P_0/2$ would now read $P_u/2$ or $P_u/2+P_p$, respectively. A straightforward calculation shows that $P_2$ would change by a factor $1+\text{DOP}$ whereas $P_1$ would change by a factor $1-\text{DOP}$ or vice-versa. In the extreme case where $\text{DOP} = 1$, it becomes clear that the measurement does not make sense: the $\xi_{R,T}$ would either be equal to zero or infinity. In a realistic scenario where $\text{DOP} \ll 1$, a first-order approximation shows that the error in the measurement, in dB, would correspond to approximately $10\cdot\log_{10}(1 + 3\cdot \text{DOP})$; for instance, if $DOP=0.1$, the error introduced in the measurement would be $\sim 1$ dB. This is in contrast with the usage of highly polarized probe lights, where the sensitivity to the DOP variation is much higher \cite{penninckx2005definition}.

It is interesting to note that, in the ideal case where the probe light is perfectly unpolarized, and assuming that the impractical pre-compensating solutions discussed in the previous Section are not employed, which is reasonable, the PDL of the devices increase the degree of polarization of the probe light. This, in turn, increases the portion of polarized light that is sent to the DUT and also affects the uncertainty of the measurement. Since the PDL is usually expressed in dB, and using the same first order approximation as above, the error due to its effect on unpolarized light, or PDL-induced DOP ($\text{DOP}_{\text{PDL}}$), can be similarly calculated to be $\text{DOP}_{\text{PDL}}= (1-10^{-\text{PDL}/10})$. Expected values of the PDL in the range $0.1-0.3$ dB, as previously stated, allow the first order approximation to hold and the introduced error is, then, $10\cdot\log_{10}(1+3\cdot\text{DOP}_{\text{PDL}})$. In the worst case scenario of a 0.3 dB PDL, this amounts to a $\sim 0.8$ dB error in the measurement. 

\subsection{Broadband vs. Laser-line PBS}
\label{sec:coating}

A PBS is usually anti-reflection coated, with the coating bandwidth depending on the application. Whereas a broadband PBS may accept a wide range of input wavelengths, this is not the case for laser-line PBS devices where a narrow linewidth is expected.

In this case, the direct use of a LED may introduce errors, as the usually negligible return loss can be decreased by orders of magnitude. A simple solution can be employed in this scenario: the ULS in Fig. \ref{fig:scheme} can be comprised of a LED, an optical circulator, and a Fiber Bragg Grating (FBG) with center wavelength coinciding with the PBS AR-coating reflection peak and bandwidth equal to or lower than the coating bandwidth.

\subsection{Applications to other devices}
\label{sec:other_devices}

The reader may have noticed that all equations (\ref{eq:3}-\ref{eq:8}) could also be written for a generic 3-port device, such as a regular beamsplitter. However, as discussed in Section \ref{sec:pdl}, there is an uncertainty in the measurement introduced by the PDL of the circulator and the optical switch, which are unavoidable. Typically, regular beamsplitters have the same PDL levels of circulators and switches, so the uncertainty introduced by these components could obscure the measurement.

\section{Conclusion}

A new "plug-and-play" method for characterization of polarizing beamsplitters that completely avoids polarization alignment procedures has been proposed and experimentally verified. The insertion losses and extinction ratios of 3- 4-port devices have been calculated and experimentally validated. The accuracy of the method is limited by the PDL of the optical circulator and optical switch as well as by the DOP of the light source, which were both shown to be upper bounded by 1 dB.

\section*{Funding}
This work was supported by CAPES, CNPq and FAPERJ.

\section*{Acknowledgments}
The authors acknowledge contributions in the early forms of the experimental setup from Marcelo Resende and Henrique Saraiva.

\section*{Disclosures}
The authors declare that there are no conflicts of interest related to this article.

\bibliographystyle{IEEEtran}
\bibliography{sample}

\end{document}